# Predictability of warm and cold winters: Assessment of El Niño effects in the North Eurasian regions

I.I. Mokhov[1,2]
[1]A.M. Obukhov Institute of Atmospheric Physics RAS
[2]Lomonosov Moscow State University
mokhov@ifaran.ru

**Abstract**
Frequency of warm and cold winters in the North Eurasian regions is analyzed from long-term data, depending on El Nino phenomena of different types. Frequencies of extremely warm and extremely cold winters for North Eurasian regions in different phases of El Niño phenomena are compared. Potential predictability of anomalous winters in the El Niño, La Niña and neutral phases is estimated.

**Introduction**

The last decades have been characterized by significant global and regional climatic changes, especially at high latitudes [1-3] (see also [4]). Against the background of general warming, there is an increase in climatic variability - "nervousness" of the climate. The most noticeable new tendencies manifest themselves in the frequency and intensity of extreme regional phenomena and processes. An increase in global surface temperature is accompanied by an increase in the number of weather and climatic anomalies. In Russia, the number of dangerous meteorological phenomena over the past two decades has approximately tripled against the background of rapid warming in Russia - at a rate of about 0.5 K per decade (two and a half times faster than global), and in a number of regions - about 1 K per decade (http://www.meteorf.ru). Since the end of the 20th century, the number of dangerous meteorological events in Russian regions has increased by an average of more than two dozen events per year [5].

The formation of seasonal anomalies is related to a series of key processes on the regional and global scale [1-3]. The formation of weather and climate anomalies in the middle latitudes of the Northern Hemisphere depends on the zonal transport in the troposphere from west to east, which is determined by the pressure differential between the subtropical and subpolar latitudes. The corresponding North Atlantic Oscillation (NAO) index is of key importance for the North Eurasia regions, in particular. It is characterized by the pressure differential between two atmospheric action centers in North Atlantic—the Azores High and the Icelandic Low [6-8]. The record warm winter of 2019–2020 was associated with a deepening of the Icelandic Low and strengthening of the Azores High with large values of the NAO index and intense atmospheric transport from the warmer Atlantic Ocean to colder continental regions. Also, the record-warm temperature of the upper oceanic layer in the Atlantic Ocean was reached in the winter of 2019/2020 according to long-term data (https://www.nodc.noaa.gov/). In recent years, the increase in the sea surface temperature (SST) in the positive phase of the Atlantic Multidecadal Oscillation (AMO) with a period of about six decades (https://www.esrl.noaa.gov) related to the Atlantic thermohaline circulation was added to the general secular trend of warming [9]. Along with this, the formation of the anomalously warm winter of 2019/2020 was favored by the fact that the influence of the Siberian High—a powerful atmospheric action center in the path of the zonal flow in the midlatitude troposphere—was weaker than usual in winter. According to model estimates, a general weakening of the winter Siberian High is expected with further global warming; this is already manifested in the long-period data [10].

The extreme regional weather and climate anomalies in the middle latitudes within a season are related to the irregular course of the atmospheric jet flow and the formation of quasi-stationary planetary waves and atmospheric blockings of the zonal flow [7,8]. In contrast to the winter of 2019/2020, regimes are formed with alternation of regions with temperature anomalies of different signs in the middle latitude belt, with meridional flows of cold air from the polar latitudes and warm air from the lower latitudes. According to model estimates [8,11] one should expect an increase in the repeatability of atmospheric blockings and meridional penetrations, with increased risk of severe frosts in winter and anomalous heat in summer under continuing global warming. The severity of weather and climate anomalies and their frequency in the middle latitudes also depend on climatic conditions in the Arctic, particularly on the Arctic atmospheric center of action [10,12]. Analysis of long-period data and the results of model simulations testify about the general trend of Arctic anticyclone weakening in the process of global warming [10].

Interannual variations in the climatic system on the whole, in particular, interannual variability of the global surface temperature, are most significantly influenced by El Nino processes. Their influence is manifested in different regions from the equatorial latitudes to the polar ones, including those of Northern Eurasia [13-21]. The relation to El Nino phenomena is also manifested for regimes of atmospheric blockings in the middle latitudes [7,22,23] and even for processes in the Arctic [10].

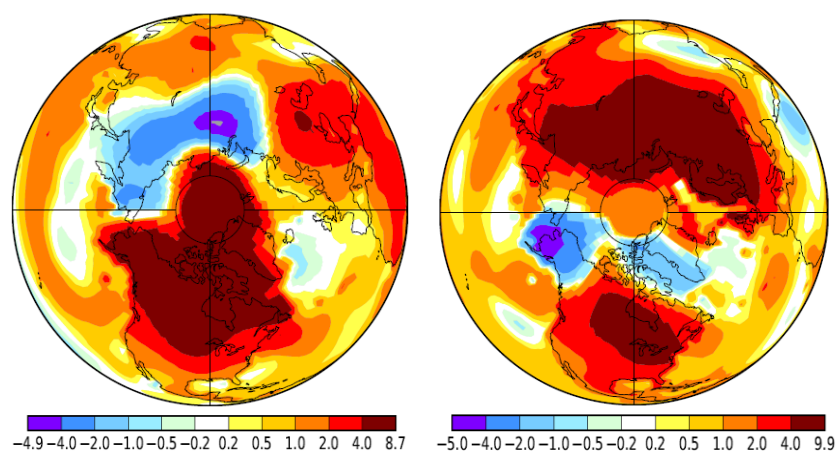

Figure 1. Regional SAT anomalies (K) in January 2021 (left) and in January 2020 (right) from GISS data.

In recent years, very contrasting regional weather and climatic anomalies have appeared. Figure 1 shows strong regional anomalies of the surface air temperature (SAT) in January 2021 and in January 2020 from GISS data (https://data.giss.nasa.gov/gistemp/). This paper presents some estimates of the predictability of anomalous winters in the North Eurasian regions at different phases of El Niño events based on long-term data.

**Data analyzed**

Here, the frequency of warm and cold winters, including extremely warm and extremely cold winters, in the North Eurasian regions is estimated for different phases of El Niño. This paper presents estimates for mid-latitude and southern regions in the European and Asian parts of Russia. The analysis involved the characteristics of warm and cold winters based on monthly averaged data for anomalies of the surface air temperature $\delta T$ in January and February for different Russian regions from meteorological observations for the period of 1936–2014 [24]. The ratio of temperature anomalies $\delta T$ in January and February to the standard deviation σ*T* for the period of 1961–1990, i.e., index *I* = δ*T*/σ*T*, was analyzed for different regions. Winters range from extremely warm (EW) and extremely cold (EC) to considerably warm (CW) and considerably cold (CC), as well as moderately warm (MW) and moderately cold (MC) winters. To the south of 60°N, EC-winters were characterized by indices *I* less than –0.9 for the European part of Russia (ER), as well as for the Amur River region and Primorye, and less than –1 for Cisbaikalia and Transbaikalia. At values of the index *I* between –0.5 and –0.9, winters were characterized as CC; at values of *I* between –0.5 and 0, as MC. Correspondingly, to the south of 60°N, EW-winters were characterized by indices *I* larger than 1.0 for the European part of Russia (ER), as well as for the Amur River region and Primorye, and for Cisbaikalia and Transbaikalia. At values of the index *I* between 0.5 and 1.0, winters were characterized as CW; at values of *I* between 0.5 and 0, as MW.

The regional effects of El-Nino phenomena were analyzed with allowance for different manifestations of El-Nino processes. Along with canonical El-Nino manifestations with positive anomalies of the sea surface temperature (SST) in the eastern part of the Pacific in the equatorial latitudes, El-Nino events with the largest SST anomalies of equatorial regions are manifested increasingly frequently in the central part of the Pacific (e.g., in [25]). The effects of El-Nino / La-Nina were estimated using their different indices characterized by the SST, in particular, in the Nino3 (150°–90°W) and Nino4 (160°E–150°W) regions in subequatorial latitudes of the Pacific (ftp://www.coaps.fsu.edu/pub/). The index in the Nino4 region characterizes the development of the El-Nino of the CP-type with considerable positive SST anomalies in the central equatorial part of the Pacific (Central Pacific, CP), in contrast to El-Nino manifestations in the eastern equatorial part of the Pacific (Eastern Pacific, EP) - El-Nino of the EP-type characterized by the index in the Nino3 region. Note that El-Nino of the CP-type is increasingly frequently observed in recent years. Phases of El-Nino (*E*) and La-Nina (*L*) were distinguished using five-month moving averaging of values of the SST anomaly. The El-Nino phase (warm phase) and La-Nina phase (cold phase) were determined by the values of SST anomalies not less than 0.5°C or not greater than –0.5°C over six months in succession, respectively. Other cases were characterized as the neutral phase (*N*).

Figure 2 shows the location of meteorological stations, the data of which were used to determine the index of winter anomalies for the selected 10 regions. Northern regions: 1 - European part of Russia (NER); 3 - Western Siberia (NWS); 5 - Central Siberia (NCS); 7 - Eastern Siberia (NES). Southern regions: 2 - European part of Russia (SER); 4 - Western Siberia (SWS); 6 - Central Siberia (SCS); 9 – Lake Baikal basin and Transbaikalia (SBT); 8 - Eastern Siberia (SES); 10 - Amur River basin and Primorye (SAP) (see [24]). The dotted line marks the border at 60º N between northern and southern regions.

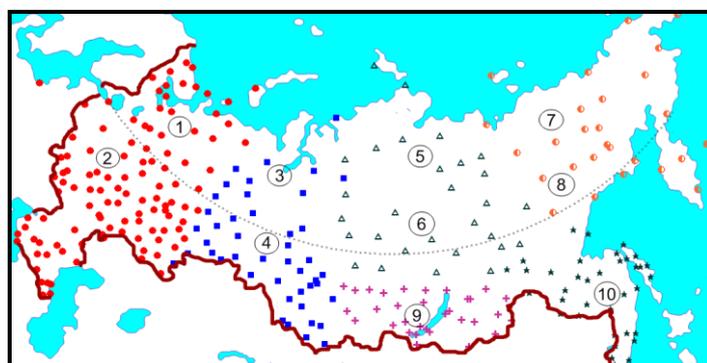

Fig. 2. Location of meteorological stations whose data were used in [Meshcherskaya, Golod 2015] to analyze winter temperature anomalies in ten selected regions. The dotted line marks the border between northern and southern regions at 60º N.

**Results**

Table 1 (a, b, c) presents the number and frequncy estimates of warm and cold winters for SER (a), SBT (b) and APR (c) south of 60°N during the onset of various El-Nino phases, characterized by Nino3 and Nino4 indices. The largest total number of cold and warm winters is characteristic for neutral (*N*) phases for all considered regions. This is due to the fact that the total number of years in the *N*-phase is greater than the total number of years in the El-Nino (*E*) and La-Nina (*L*) phases. Estimates of the frequency of an abnormal winter for the expected *E*-phase or *L*-phase can be significantly higher than the corresponding estimates for the *N*-phase. For example, the probability of a warm winter for SER is estimated to be the highest in the *E*-phase (more than 2/3), and the frequency of a cold winter is estimated as the highest in the *L*-phase.

Table 1. The number and estimates of the frequency (in brackets) of extremely warm (EW), considerably warm (CW) and moderately warm (MW), as well as extremely cold (EC), considerably cold (CC) and moderately cold (MC) winters for the SER (a), SBT (b) and APR (c) to the south from 60°N at different El Niño phases, characterized by the Nino3 and Nino4 indices.

| (a) 1936-2014 | | SER | | | | | |
|---|---|---|---|---|---|---|---|
| | | Warm Winters | | | Cold Winters | | |
| | | EW | EW+CW | EW+CW+MW | EC | EC+CC | EC+CC+MC |
| Nino3 | *N* $n_\Sigma=44$ | 4 (0.09) | 14 (0.32) | 25 (0.57) | 5 (0.11) | 11 (0.25) | 19 (0.43) |
| | *L* $n_\Sigma=19$ | 3 (0.16) | 7 (0.37) | 9 (0.47) | 3 (0.16) | 7 (0.37) | 10 (0.53) |
| | *E* $n_\Sigma=16$ | 1 (0.06) | 6 (0.31) | 11 (0.69) | 0 (0) | 1 (0.06) | 5 (0.31) |
| Nino4 | *N* $n_\Sigma=40$ | 3 (0.08) | 13 (0.33) | 21 (0.53) | 3 (0.08) | 9 (0.23) | 19 (0.48) |
| | *L* $n_\Sigma=18$ | 3 (0.17) | 7 (0.39) | 9 (0.50) | 3 (0.17) | 7 (0.39) | 9 (0.50) |
| | *E* $n_\Sigma=21$ | 2 (0.10) | 6 (0.29) | 15 (0.71) | 2 (0.10) | 3 (0.15) | 6 (0.29) |

According to Table 1a the most significant difference in the frequency of cold and warm winters in SER was noted for E-phases, and the smallest differences - for L-phases. The frequency of warm winters in the SER during the E-phases is more than twice the frequency of cold ones. The highest frequency of warm winters for SER in years starting in the *E*-phase is mainly associated with MW-winters.

The frequency of warm winters for SBT (Table 1b) is generally greater in the *L*-phase, and the frequency of cold winters is estimated to be the highest in the *E*-phase (especially with the use of index Nino3). It is important to note that, according to data for eight decades, EC-winters and CC-winters have not been detected in the L-phases in SBT.

| (b) 1936-2014 | | SBT | | | | | |
|---|---|---|---|---|---|---|---|
| | | Warm Winters | | | Cold Winters | | |
| | | EW | EW+CW | EW+CW+MW | EC | EC+CC | EC+CC+MC |
| Nino3 | $N$ $n_\Sigma=44$ | 4 (0.09) | 17 (0.39) | 22 (0.50) | 5 (0.11) | 6 (0.14) | 22 (0.25) |
| | $L$ $n_\Sigma=19$ | 4 (0.21) | 8 (0.42) | 11 (0.58) | 0 (0) | 0 (0) | 8 (0.42) |
| | $E$ $n_\Sigma=16$ | 0 (0) | 1 (0.06) | 4 (0.25) | 3 (0.19) | 8 (0.50) | 12 (0.75) |
| Nino4 | $N$ $n_\Sigma=40$ | 3 (0.08) | 13 (0.33) | 16 (0.40) | 7 (0.18) | 13 (0.33) | 24 (0.60) |
| | $L$ $n_\Sigma=18$ | 3 (0.17) | 8 (0.45) | 12 (0.67) | 0 (0) | 0 (0) | 6 (0.33) |
| | $E$ $n_\Sigma=21$ | 2 (0.10) | 3 (0.24) | 9 (0.43) | 1 (0.05) | 6 (0.26) | 12 (0.57) |

In APR (Table 1c), the highest frequency of EW- and CW-winters is in L-phases. The highest frequency of EC- and CC-winters was detected in E-phases. It is worth to note, that for APR, noticeable differences in the frequency estimates were exhibited when using different El Niño indices.

| (c) 1936-2014 | | APR | | | | | |
|---|---|---|---|---|---|---|---|
| | | Warm Winters | | | Cold Winters | | |
| | | EW | EW+CW | EW+CW+MW | EC | EC+CC | EC+CC+MC |
| Nino3 | $N$ $n_\Sigma=44$ | 5 (0.11) | 13 (0.29) | 22 (0.50) | 4 (0.09) | 11 (0.25) | 22 (0.50) |
| | $L$ $n_\Sigma=19$ | 2 (0.11) | 10 (0.53) | 11 (0.58) | 1 (0.05) | 3 (0.16) | 8 (0.42) |
| | $E$ $n_\Sigma=16$ | 1 (0.06) | 3 (0.19) | 7 (0.44) | 3 (0.19) | 5 (0.32) | 9 (0.56) |
| Nino4 | $N$ $n_\Sigma=40$ | 3 (0.08) | 10 (0.25) | 17 (0.43) | 2 (0.05) | 10 (0.25) | 23 (0.58) |
| | $L$ $n_\Sigma=18$ | 2 (0.11) | 9 (0.50) | 11 (0.61) | 2 (0.11) | 3 (0.17) | 7 (0.39) |
| | $E$ $n_\Sigma=21$ | 3 (0.14) | 7 (0.33) | 12 (0.57) | 4 (0.19) | 6 (0.29) | 9 (0.43) |

**Discussion and conclusions**

The presented estimates and the corresponding approach have predictive value and can be used to assess regional seasonal risks in connection with the development and inter-seasonal and interannual changes of key El Niño-type processes with significant global climatic influence. For SER, in particular, in E-phase, the frequency of a warm winter is estimated to be significantly (more than twice) higher than the frequency of a cold winter. At the same time, the frequency of cold winters in E-phase in SBT is much higher than in warm winters. For SBT, the manifestation of extremely and considerably cold winters (EC-winters and CC-winters) in the L-phase is unlikely - this has not been observed for eight decades.

The degree of justification of estimates obtained from the data for the period 1936-2014 can be tested for subsequent years, in particular for the winter months of 2021, which started in the L-phase, in which no extremely and considerably cold winter conditions (EC-winters and CC-winters) were identified for SBT.

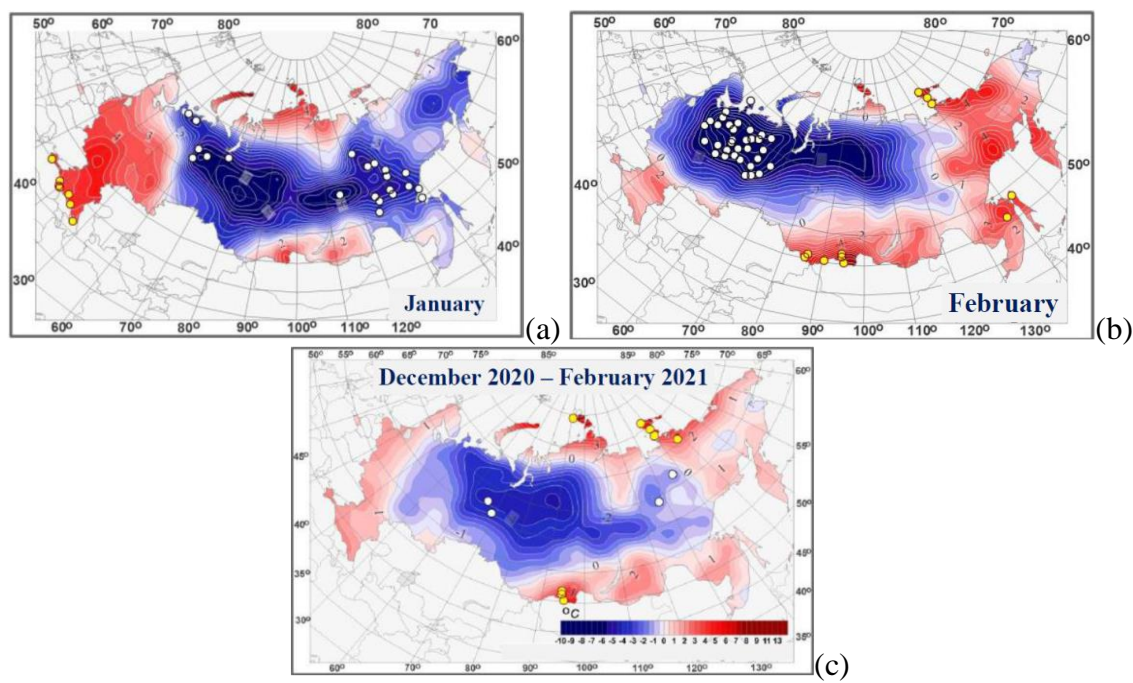

Fig. 3. Seasonal and monthly mean surface air temperature anomalies (ºC relative to 1961-1990) in Russian regions according to Roshydromet data (https://www.meteorf.ru) for winter 2020/2021 (white circles show the meteorological station location with extrema below the 5th percentile, yellow - above the 95th percentile).

Figure 3 shows seasonal and monthly mean surface air temperature anomalies (relative to 1961-1990) in Russian regions according to Roshydromet data (https://www.meteorf.ru) for winter 2020/2021. According to Fig. 3 over a large territory of Russia in the winter of 2020/2021, including in January and February 2021, significant negative SAT anomalies were revealed, while positive temperature anomalies were noted for the Lake Baikal basin and Transbaikalia. According to meteorological observations (climatechange.igce.ru), positive SAT anomalies (relative to 1961-1990) were observed for all winter months in the Lake Baikal basin and Transbaikalia. At the same time, in other large regions of Russia (including in the European and Asian parts of Russia, in Western, Central and Eastern Siberia, Amur River basin and Primorye) and in Russia as a whole, temperature anomalies were negative at least in one of the winter months, including January and February 2021. After 2014, another year, along with 2021, began in the L-phase - 2018. For the winter months of 2017/2018 according to meteorological observations, positive SAT anomalies were also observed in the Lake Baikal basin and Transbaikalia for all winter months.

Thus, the data for the last winters confirm the significance of the estimates obtained from the previously obtained data, in particular, for winters in the Lake Baikal basin and Transbaikalia in L-phase. In connection with the newly developing L-phase at the end of

2021, the frequency of which in the winter months of early 2022 according to ensemble model estimates according to CPC / IRI official probabilistic ENSO forecasts exceeds 90%, one should expect, in particular, the absence of extreme cold weather in January-February 2022 in the Lake Baikal basin and Transbaikalia.

Obtaining more reliable corresponding estimates of the regional weather and climatic predictability, indeed, requires a more detailed and comprehensive analysis of data and ensemble model simulations. Nevertheless, even the comparatively small statistical base of available data and different types of El-Nino phenomena develops effects significant for obtaining prognostic estimates of regional weather and climatic anomalies.

It is significant that key modes of climatic variability, including El Nino phenomena, and their influence on other regions change under global climate changes. In particular, the trend of intensification and increasing frequency of El-Nino phenomena with global warming, as mentioned in [26,27], is associated with an increase in the risk of their stronger influence on other regions including regions of Northern Eurasia. One should also mention the necessity of estimating the combined effects of key modes of climatic variability along with El Nino (including NAO, AMO, Pacific Decadal Oscillation).

**Acknowledgements**

This work was supported by the Russian Science Foundation (project No. 19-17-00240).